\documentclass[showpacs,twocolumn,amsmath,superscriptaddress,floatfix,pre]{revtex4}
\usepackage{epsfig,verbatim}

\begin{document}

\title{Random  Field  Potts   model  with  dipolar-like  interactions:
hysteresis, avalanches and microstructure}

\author{Benedetta Cerruti}

\email{benedetta@ecm.ub.es}

\affiliation{ Departament d'Estructura i Constituents de la Mat\`eria,
  Universitat de  Barcelona \\ Diagonal 647,  Facultat de F\'{\i}sica,
   08028 Barcelona, Catalonia}

\author{Eduard Vives}

\affiliation{ Departament d'Estructura i Constituents de la Mat\`eria,
  Universitat de  Barcelona \\ Diagonal 647,  Facultat de F\'{\i}sica,
  08028 Barcelona, Catalonia}

\date{\today}

\begin{abstract}
 A model for  the study of hysteresis and  avalanches in a first-order
 phase transition from a single  variant phase to a multivariant phase
 is  presented. The model  is based  on a  modification of  the Random
 Field  Potts model  with  metastable dynamics by  adding a  dipolar
 interaction term truncated at  nearest neighbors. We focus our study
 on  hysteresis  loop  properties,  on   the  three-dimensional
 (3D) microstructure formation and on avalanche statistics.
\end{abstract}

\pacs{75.60.Ej, 75.50.Lk, 81.30.Kf, 64.60.Cn}

\maketitle

\section{Introduction}
\label{Introduction}

Microstructure  formation  in   first-order  phase  transitions  is  a
phenomenon  that has  been studied  by physicists,  mathematicians and
engineers \cite{Salje1990,Bhattacharya2003,Khachaturyan}.   
It is  important not only  from a
fundamental  point of  view,  but  also for  applications,  due to  its
relationship  with material  properties. Microstructures  occur  in ferroic
systems (ferromagnetic,  ferroelectric and/or ferroelastic)  which are
driven through  a first-order phase transition  (FOPT) in  which some
symmetry operations of  the parent phase are lost.   The product phase
(usually  less  symmetric)  may  appear  in  energetically  equivalent
variants which are  related by the symmetry operations  that have been
lost in  the transition.   The obtained microstructures  correspond to
the arrangement of such equivalent  variants and are decided by the
interplay of  many energetic  terms: interface energy,  surface energy
and long-range interactions.

Until now many of the  studies on microstructures have focused on the
determination of  the optimal  variant configuration minimizing a
certain  thermodynamic potential  that  takes into  account the  above
factors and external conditions. Nevertheless, in many cases, when
real  materials  are studied,  such  optimal  microstructures are  not
observed.   This is  mainly  due  to two  important  factors: (i)  the
existence of disorder sources  of very different natures both in the
bulk and on  the surfaces and also (ii) the  athermal character of the
phase transition dynamics: when the  temperature is not very high, the
energetic barriers that separate the optimal solutions from the parent
phase cannot  be overcome. Thus, the system  evolves following metastable
paths which  locally optimize the system  energy, but are  far from the
trajectories  obtained  from  a  global  minimization  principle.   An
interesting  suggestion on  the behavior  of  microstructure formation
comes  from  the  glass-jamming   transition  framework  (see  Refs.
\onlinecite{Toninelli2006,  Toninelli2006b}  and references  therein),
which  is associated with the so-called  kinetically constrained model.
These models  are stochastic lattice  gases with hard  core exclusion,
with the addition  of some local constraints, which  mimic the geometric
constraints  on the  possible  rearrangements in  physical systems. 
Similar behavior is quite likely to arise in microstructure formation.

The use of continuum models derived from  elasticity theory
\cite{Ahluwalia2001,Jacobs,Hatch2003,Lookman2003,Rasmussen2001} has been
proposed as another approach to microstructure formation.    Some   of
these  models  have  been  successful in  explaining  microstructures,
hysteresis and avalanches. Nevertheless, they are very time consuming
from a computational point of view. For this reason, in many cases
only 2D problems have been addressed, and even this item presents
difficulties connected with large statistics  and with the scanning of 
the  model parameters in  order to study their influence.
Our aim here is to find a statistical mechanical lattice model,
easy to simulate and which allows for the study of statistics of 
microstructure sequences dynamically generated in athermal systems, under the
influence of disorder.

The Random-Field Ising  model (RFIM) \cite{Sethna1993} with metastable
dynamics is one  of the simplest models for the study of the combined effects
of disorder  and athermal  evolution. It is  formulated in  a magnetic
language  for  a  spin  reversal  transition, driven  by  an  external
magnetic field  $H$. The  only degrees of  freedom are  spin variables
defined on a lattice $(i=1,...,N)$, 
which take values $S_i=\pm 1$ on the $i$-th lattice
site. The RFIM enables computation of
hysteresis loops $M(H)$  corresponding to the  behavior of
the order parameter $M=\sum_i S_i$ as a function of $H$ as well as the
analysis of  the intermediate states between  the negatively saturated
initial  state $\{S_i=-1\}$ 
and  the final  positively saturated  state
$\{S_i=+1\}$, and  {\it vice versa}. 
In  particular, the RFIM  has been
successful in  understanding the avalanche  dynamics (Barkhausen noise
\cite{Barkhausen1919}) that joins  the intermediate metastable states
and shows  absence of characteristic  scales for a critical  amount of
disorder.   Moreover,  the  RFIM  displays several  other  interesting
properties \cite{Sethna1993,Dhar1997}: it exhibits a well-defined 
rate-independent trajectory,  it shows  return point memory, it satisfies the
abelian property and,  from a computational point of  view, it is fast
to simulate trajectories in relatively large systems \cite{Kuntz1999}.

Nevertheless,   the  usefulness  of   the  RFIM   for  the   study  of
microstructures is almost  null. This happens because the
parent and product phases in the RFIM 
are  the positively magnetized phase and the
negatively magnetized phase.  These  two phases are single variant and
totally equivalent  from a symmetry point of  view.  Consequently, the
obtained  hysteresis   loops  are   symmetric  under  the  exchange  $H
\rightarrow -H$ and $M\rightarrow -M$, there is absence of latent heat
associated  with the  FOPT  and the  domains  are spherically  symmetric
(except  for  some  short-range  correlations  due  to  lattice
symmetries).

Within this  framework, the  aim of  the present work  is to  explore some
modifications that  should be introduced in  the RFIM in  order to obtain
3D  microstructures without losing, as far  as possible, some
of the useful RFIM properties that we have mentioned above.  A first step
in this direction was done  several years ago by defining the Random
Field   Blume-Emery-Griffiths   model, in which the 'spin' variables take
three different values ($S_i=-1,0,1$)   with  metastable   dynamics
\cite{Vives1995}.  In  this case, the  FOPT takes place from  a single
variant parent phase, represented by $S_i=0$, and a  product phase with
two  variants  $S_i=+1$  and  $S_i=-1$. In the cited work, 
the  hysteresis  loops,  phase
diagram  and avalanche  distribution were  studied for this type of simple
case.
In the present paper we would like to go one step forward. This will be done
by starting from the  Random Field Potts model \cite{Blankschtein1984}
with metastable dynamics. In the Potts model the spin variables can take
an arbitrary number of values. This model allows phase transitions to be
studied from  a  non-degenerate  phase  $S_i=0$  to  a  multivariant  product
phase. We will explore the  effect of an extra interaction 
term  (of a dipolar nature,  but truncated  to the nearest-neighbor
approximation), which
will be necessary in order  to produce microstructures (for the introduction
of dipolar interactions in RFIM, see \cite{magni}).  It is not our
aim  to   focus  on  the  detailed  analysis   of  any  particular
transition.  We will study a model that, from the point of view of
symmetry,  would  correspond to  a  transition  from  a cubic  phase
(single  variant)  to  a   tetragonal  phase  (with  three  equivalent
variants), although neither the detailed interactions nor the external
constraints  will  be  tuned  for  the  particular  modeling  of  such
transitions in ferroelastic systems.  This will be the aim of a future
work \footnote{B. Cerruti and E. Vives, {\it in preparation}}.

The paper is organized as follows: in section \ref{Model} we introduce
the hamiltonian of the model.  In section \ref{Dynamics} we detail the
metastable dynamics  that has been used for  the simulations.  Section
\ref{Results} is devoted to the discussion of the obtained results: in
section  \ref{area_asimmetria} we present our  analysis on  the
shape of hysteresis  loop cycles as a function  of the model parameter
values.  In this section it is shown that the 
loops happen to  be unsymmetric, in
contrast to  the RFIM results.  The microstructures  are analyzed in
section  \ref{Microstructures},   where  we  discuss three different
regimes corresponding to different ranges of the parameters values.
Moreover,  in  section  \ref{Avalanches}  we present  the statistical
analysis of avalanche behavior.  Finally, we summarize and discuss the
future perspectives in section \ref{Conclusions}.

\section{Model}
\label{Model}
The model  can be defined on  any regular lattice. We  will consider a
simple  cubic lattice  of  size $N=L\times  L\times  L$ with  periodic
boundary conditions. At  each lattice site we define  a variable $S_i$
($i=1,\dots N$), which can take  four different values that we will call
$0$, $x$,  $y$ and $z$.   We can choose different  representations for
our variables  but it is  convenient to consider a  vector $\vec{S_i}$
having three components: we will  indicate the four possible values as
$0=(0,0,0)$, $x=(1,0,0)$, $y=(0,1,0)$ and $z=(0,0,1)$.

The  order parameter  for the  phase  transition under  study here  is
$M=\sum_i  (\vec{S}_i)^2$,  where  the  sum  spans  over  the  whole
lattice. $M$ represents the amount of the system that has transformed
from the cubic 
to the tetragonal  phase.  By following the analogy  with the magnetic
case, we will refer to  $M$ as the total magnetization of
the system.  Moreover, we define the normalized magnetization $m$,
as $m=M/N$. We will drive the system by an external field $H$ coupled
to $M$, since we are interested in the transition  from the $0$ phase
to the multivariant phase which will be composed of regions (variants)
in the states $x$, $y$ and $z$. The field $H$ would correspond to
the driving effect of the temperature in athermal structural transitions.
We will start by decreasing  $H$, from the $M=0$ state.
We consider the following hamiltonian:
\begin{multline}
{\cal H } =  -k \sum_{<ij>}^{NN} \delta(\vec{S}_i,\vec{S}_j) + \lambda
\sum_{<ij>}^{NN} \frac{(\vec{S}_i  \cdot \vec{r}_{ij})(\vec{S}_j \cdot
\vec{r}_{ij})}{|\vec{r}_{ij}|^3}            \\           +           H
\sum_i^N(\vec{S}_i)^2+{\cal{H}}_{dis}
\label{hamiltoniana}
\end{multline}
The first term is a Potts exchange term extending to nearest-neighbor
(n.n.) pairs. The parameter $k$ will be always positive, favoring the
ferromagnetic  interaction between spins  in the  same state.   In the
following, without loss of generality, we will consider $k=1$.

The  second   term  is a  dipolar interaction  truncated  to  n.n.
pairs. As we have already said, the aim of adding this term 
is to generate some simple
microstructures.   To include higher  order terms  would lead  to more
realistic  ones.  The  vector ${\vec  r}_{ij}$ is  the  lattice vector
joining 'spins' ${\vec S}_i$ and  ${\vec S}_j$. We will study 
the cases  with $\lambda<0$  and $\lambda>0$ separately.  
As  can be  easily seen
from Eq. \ref{hamiltoniana}, in fact, the $\lambda<0$ case corresponds
to favoring the  growth of prolate (needle-like)  domains parallel to
the spin direction. On the other hand, 
for $\lambda>0$  such a growth is not favored, but
as it is partially  compensated  by   the  exchange  term  it  basically
corresponds  to   the  formation   of  oblate  (disk-like)  domains,
perpendicular  to  the  spin  direction.   We  will  illustrate  these
features in Sec.  \ref{Microstructures}.

The third  term of the hamiltonian 
accounts  for the interaction between the system and the external field
$H$.  This field will be driven from very high positive values to very
negative  values (and vice versa)  in steps  $\Delta H$,  mimicking an
adiabatic triangular driving force (i.e.  field frequency $\omega\rightarrow
0$). By a deliberate abuse of language  we will refer to the step $\Delta
H$ as the driving rate. One can notice that it is possible to add 
a second driving term $\vec{G}\sum \vec{S}_i$ which would be convenient for the
study of the transitions from one variant to another, mimicking, for instance,
the effect of an applied external stress.

The last term ${\cal{H}}_{dis}$  accounts for the quenched disorder of
the system.  We will restrict ourselves to a random field type with zero
averages. However,  there are  still several  possibilities for
such a hamiltonian  term because the random fields  can couple either
to $\vec{S}_i$ or to the order parameter $(\vec{S}_i)^2$. We will thus
consider:
\begin{equation}
{\cal{H}}_{dis}=\sigma\sum_i^N\vec{g}_i\cdot\vec{S}_i+
\rho\sum_i^N{h}_i(\vec{S}_i)^2
\end{equation}
where  $\vec{g}_i$ is  a three-component vector  random  field, whose
components are  extracted from  a gaussian distribution  $N(0,1)$ with
zero mean and unitary standard deviation, and ${h}_i$  is a scalar field,
again extracted  from a gaussian distribution  $N(0,1)$.  The parameters 
$\sigma$ and $\rho$ control the amount  of quenched
disorder in the system.

In order  to compare our model  with the standard RFIM,  we define the
total amount  of disorder $\sigma_0^2=\sigma^2+\rho^2$.   In practice,
the  two disorder  terms can  be understood  as arising  from  a local
random field $\vec{f}_i$, whose components are correlated, being
\begin{equation}
\vec{f}_i=\sigma(g_{ix},g_{iy},g_{iz})+\rho(h_i,h_i,h_i),
\end{equation}
so that $\langle f_{ix}^2 \rangle = \langle f_{ix}^2 \rangle = \langle
f_{ix}^2  \rangle=\sigma_0^2$  and $\langle  f_{ix}  f_{iy} \rangle  =
\langle f_{ix} f_{iz}\rangle = \langle f_{iy} f_{iz} \rangle=\rho^2$.

\section{Dynamics}
\label{Dynamics}
\begin{figure}[ht]
\begin{center}
\epsfig{file=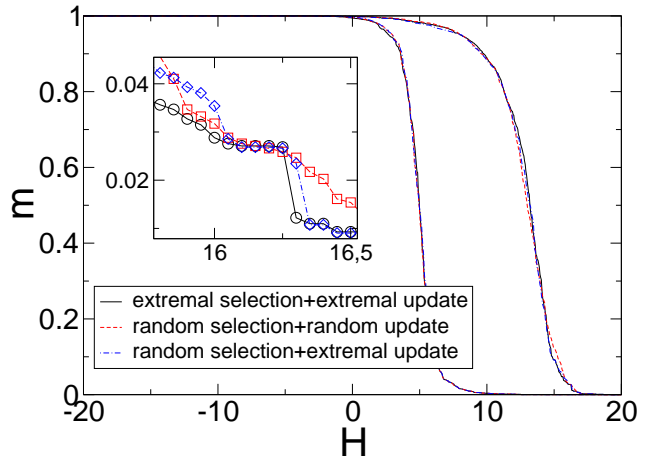,width=7.2cm,angle=-90,clip=}
\end{center}
\caption{\label{confronto_dinamiche} (Color on line) Hysteresis cycles
  for  three  different  dynamics  as  explained  in  the  text.   The
  parameters  of   the  simulations  are:   $\Delta  H=0.05$,  $L=16$,
  $\lambda=-10$, $\sigma=5$ and $\rho=2$. In the inset: a magnification of a
  loop region: the magnetization values for the three dynamics 
  coincide only for some field values.}
\end{figure}

There  are  several  possibilities  for  the choice  of  metastable
dynamics.   In Fig.~\ref{confronto_dinamiche} we  show  examples of
hysteresis  loops  obtained  with   three  possible  choices  of 
dynamics. At first sight, the three loops look very similar.  In all
the cases  we start from a  metastable state, we  increase or decrease
the  field by a $\Delta H$ step and then,  at constant  field,  we
recursively minimize the  system  energy  by  using  a  local  rule  based  on
single-spin changes. Only  after a new metastable state  is reached we
proceed with a new field change $\Delta H$.

In the  {\it extremal selection +  extremal update} case,  we scan the
whole system and check which variable  $S_i$ can change to a new value
with the  minimum energy difference  $\Delta {\cal H}$.   The proposed
change is accepted if this  minimum $\Delta {\cal H}$ is negative.  In
the {\it random selection +  random update} case, we randomly choose a
spin on  the lattice and propose a  random change to a  new value.  If
the  proposed change  implies  $\Delta {\cal  H}  < 0$  the change  is
accepted.  Finally, in the {\it  random selection +  extremal update}
dynamics, we randomly choose  a spin on the  lattice and check among
the three possible  new values which one represents  a minimum $\Delta
{\cal  H}$.   If  this  minimum   value  is  negative  we  accept  the
change. From  a computational point of  view the first  choice is much
more time consuming than the  other two, since the effort scales with
$L^3$.

Although the  loops are very  similar, detailed  analysis reveals
that  the obtained hysteresis loops,  as  well  as the  sequence  of
metastable states, are not identical.  This tells us that the proposed
model is not abelian and that the final state will depend on the order
in which  unstable spins  will be  changed. In  order to  ensure some
robustness  of the  results, we  are thus  forced to  choose {\it
extremal selection + extremal  update} dynamics, that is: to propose 
the  optimal spin change  among the whole  lattice and
among all  the three possible final  values at
each time  step. This kind  of dynamics is
deterministic  and thus,  by definition,  independent of  the updating
order.  We will keep to this dynamics for the rest of the paper.

\begin{figure}[ht]
\begin{center}
\epsfig{file=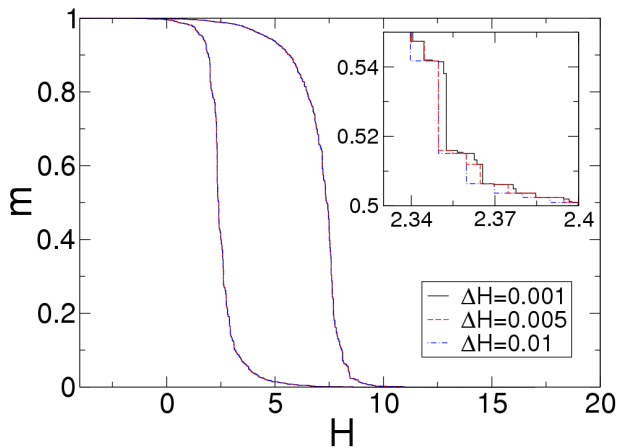,width=7cm,,angle=-90,clip=}
\end{center}
\caption{\label{DHvari} (Color on  line) Hysteresis cycles for various
  values of the  applied field rate $\Delta H$  with the {\it extremal
  selection+extremal   update}  dynamics.    The  parameters   of  the
  simulations  are: $L=16$,  $\lambda=-5$, $\sigma=4$ and
  $\rho=0$.   In  the inset:  a  magnification  of  a portion  of  the
  hysteresis cycle.}
\end{figure}

We now study   the  effect  of  changing  the value of 
$\Delta   H$.   In  Fig.
\ref{DHvari} we  show three  hysteresis cycles obtained  for different
values  of  the  driving  rate  $\Delta H$  using  the  {\it  extremal
selection+extremal update} dynamics.  A detailed analysis reveals that
the differences between the three  loops can be attributed to the fact
that driving with a smaller  $\Delta H$ allows more metastable
intermediate states to be found,  
but for  the same applied  field values,  in the
three  realizations not  only the  magnetization, but  also  the final
microscopic configurations reached are the same. The independence from
the field rate is an important property from the point of view of
the simulations,  since it allows us  to use a  relatively large $\Delta
H$ for the study of the properties of hysteresis loops.

\section{Results}
\label{Results}
We  have  performed  numerical  simulations  of systems  with  sizes
$L=8,16,32,40$ and  $60$, averaging over $10^2-10^3$  realizations of the
quenched random fields.  We  have focused our analysis on hysteresis loops
behavior,  on  microstructure  formation and  on  the  
statistical properties of the avalanches.

\subsection{Hysteresis loops}
\label{area_asimmetria}

\begin{figure}[ht]
\centerline{\epsfig{file=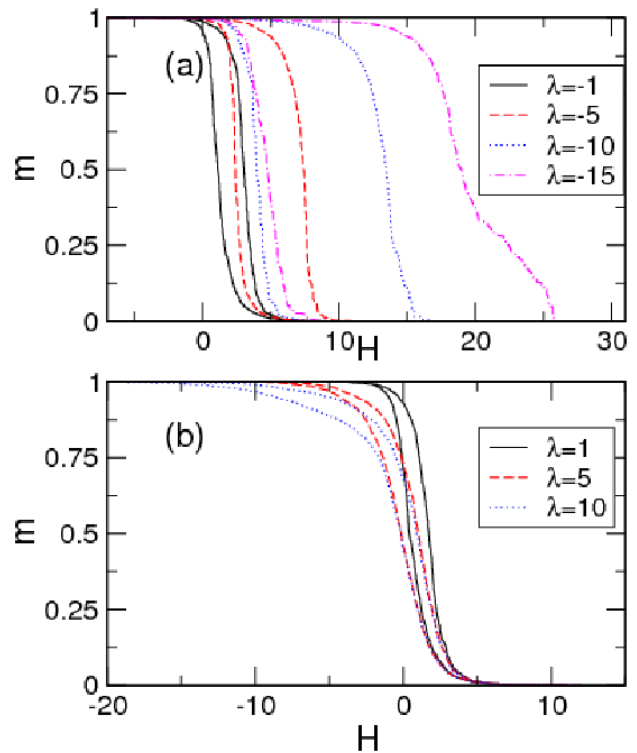,width=8.2cm,clip=}}
\caption{\label{lambda_vari}(Color  on line) Examples  of hysteresis loops  for different
values of the parameter $\lambda$: (a) $\lambda=-1,-5,-10,-15$ and (b)
$\lambda=1,5,10$.   In   all  the  cases,   $L=16$,  $\Delta  H=0.01$,
$\sigma=3$ and $\rho=0$ }
\end{figure}

\begin{figure}[ht]
\hspace{4cm}
\begin{center}
\epsfig{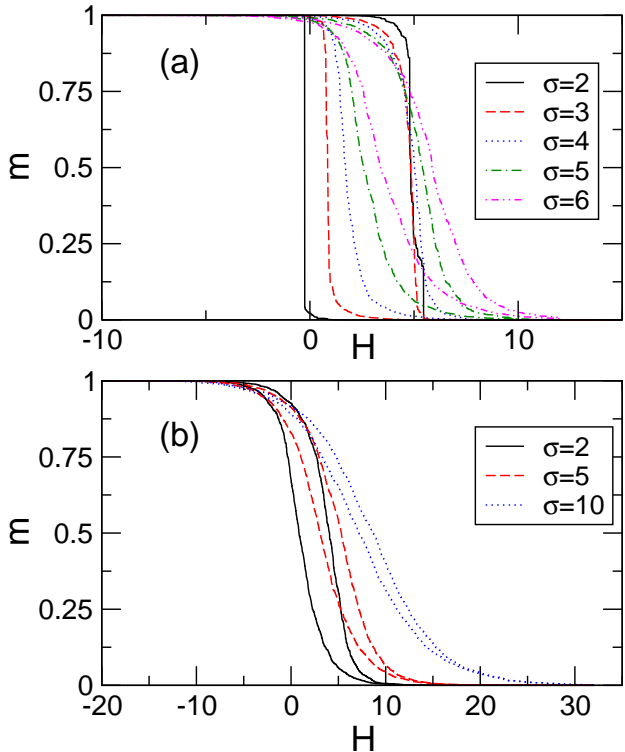}
\end{center}
\caption{\label{sigma_vari}(Color  on line)  Examples  of hysteresis  loops for  various
values  of the  disorder parameter  $\sigma$: (a)  for $\sigma=4,5,6$,
with $\rho=0$; and (b) for  $\sigma=2,4,10$ with $\rho=5$.  In all the
cases $\lambda=-3$, $L=16$ and $\Delta H=0.01$. }
\end{figure}

\begin{figure}[ht]
\centerline
\centerline{ \epsfig{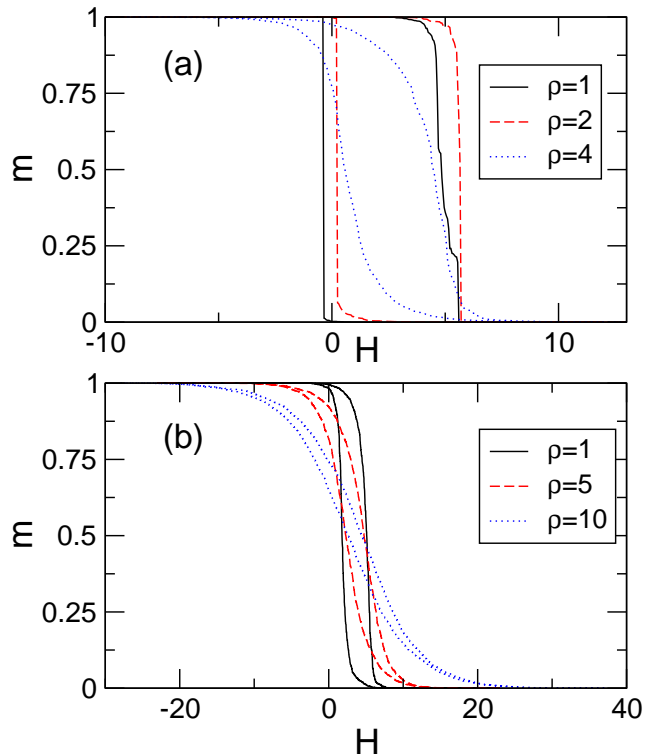}}
\caption{\label{rho_vari}(Color  on line) Examples of  hysteresis cycles  for various
values  of the disorder  parameter $\rho$:  (a) for  $\rho=1,2,4$ with
$\sigma=1.5$; and  (b) for $\rho=1,5,10$ with $\sigma=4$.   In all the
cases, $\lambda=-3$, $L=16$, and $\Delta H=0.01$.  }
\end{figure}

In  Figs.~\ref{lambda_vari},  \ref{sigma_vari} and  \ref{rho_vari}, we
show some  examples of hysteresis  loops simulated with our  model in
order  to   illustrate  the   effect  of  the   different  hamiltonian
parameters.

We   consider   the   cases   with   $\lambda<0$   
(Fig.~\ref{lambda_vari}(a))  and  $\lambda>0$   separately  
(Fig.~\ref{lambda_vari}(b)),
because they show a  clearly different  behavior.   In the  first case, which
corresponds  to the  formation of  prolate domains,  the  more negative
$\lambda$ is,  the larger the width of  the loop. In the  case of very
negative  values,  the  loops  start  to  exhibit  a  plateau  in  the
increasing field branch: the retransformation to the $0$ phase is done
in two  separate steps.  (This  effect will be discussed  below).  For
the second case, increasing lambda towards positive values increases
the tilt of the hysteresis  loop so that saturation in the transformed
phase  can only  be obtained  when the  field is  very  negative. This
effect is due to the competition between the Potts and the dipolar
terms.  Many  domains in the final stages  of the  transformation are
frustrated and  can only be transformed  by a very  negative $H$, as
occurs   in  ferromagnets   that   contain  a   small  percentage  of
antiferromagnetic bonds.

In Figs.~ \ref{sigma_vari} and \ref{rho_vari}  we show the  effect of
the two  disorder parameters $\sigma$ (Fig.~\ref{sigma_vari}, in the
low (a) and high (b)  $\rho$ regimes) and $\rho$ (Fig.~\ref{rho_vari},
in the low (a) and high (b) $\sigma$ regimes). In all the cases it can
be seen that increasing the  amount of disorder increases the tilt and
decreases the width of the loop. Moreover, as expected, for low values
of the  amount of  disorder ($\sigma$ or  $\rho$) the loops  exhibit
sharp  discontinuous  (ferromagnetic-like)  behavior.   This feature is  in
agreement with recent results on the standard RFIM, concerning the  
observation that the transition
from  sharp to  smooth  loops can  be  induced by  different kinds  of
disorder parameters:  not only the random field  variance $\sigma$ but
also  random  anisotropy \cite{Vives2001}, the  vacancy concentration
\cite{Illa2006b}, etc..  Our model  shows that the correlation with
the random fields of intensity $\rho$ can also act in a similar way.

\begin{figure}[ht]
\begin{center}
\includegraphics[width=7cm,angle=-90,clip=]{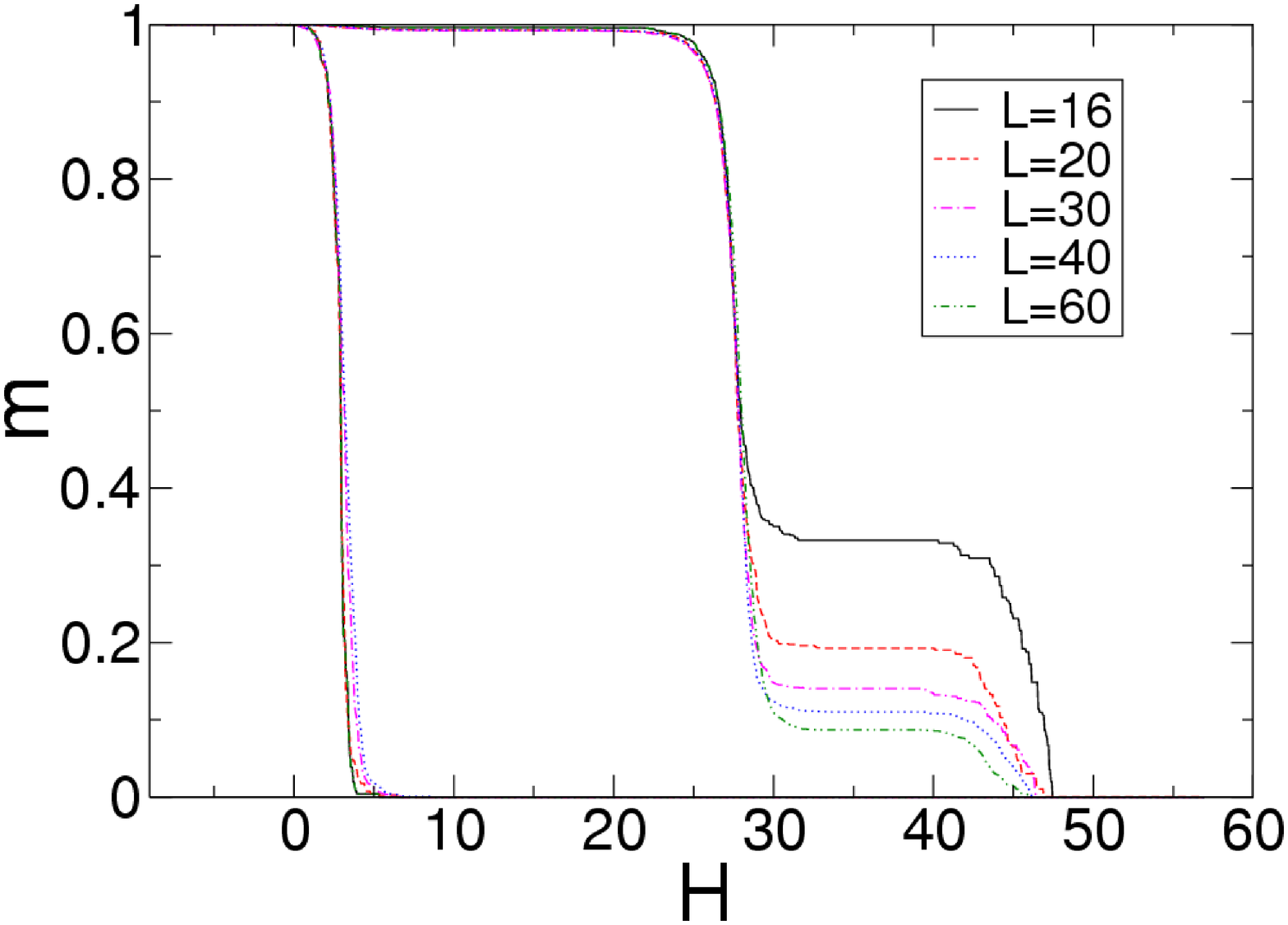}
\end{center}
\caption{\label{dimensione_variabile} (Color  on line) Hysteresis loop
  for various  values of the  system size $L=16,20,30,40,60$.   The model
  parameters  are:  $\lambda=-25$,  $\sigma=3$,  $\rho=0$  and  $\Delta
  H=0.05$}
\end{figure}

Let us now discuss  the plateau observed in Fig.~\ref{lambda_vari}(a)
in the  increasing field branch.   As shown in the  examples in
Fig.~\ref{dimensione_variabile},  this   plateau  occurs  at  smaller
magnetizations when the system size is increased. This suggest that it
may be  due to  the stabilization of  ``slab'' domains that  cross the
whole  system from  one face  to the  other and that  due to  the periodic
boundary conditions  behave as infinitely  large. Such slabs  become less
and  less  frequent by increasing the system size. This  suggestion  has  been
confirmed by analyzing sequences of configurations. An example will be
discussed in Sec. \ref{Microstructures}.

\begin{figure}[ht]
\begin{center}
\includegraphics[width=7cm,angle=-90,clip=]{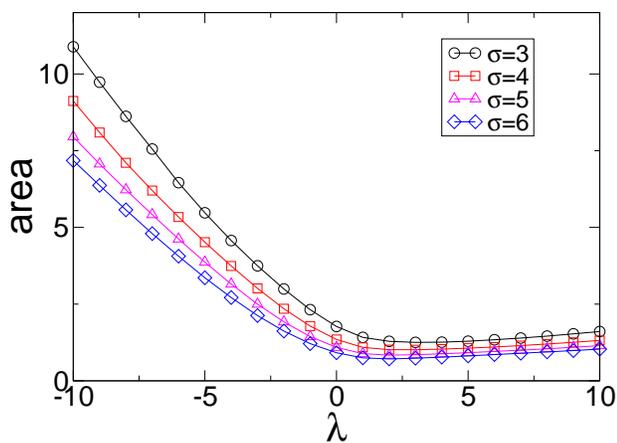}
\end{center}
\caption{\label{area}  (Color  on  line)  Hysteresis loop  area  as  a
function  of  the  parameters  $\lambda$, for  $\sigma=3,4,5,6$.   The
system  parameters are:  $L=16$, $\Delta  H=0.05$ and  $\rho=0$.  Each
point represents  an average over $800$ realizations  of the disorder.
Solid lines are a guide to the eye.  Error bars  are not visible on
the scale of the picture.}
\end{figure}

In order to perform a quantitative analysis of the hysteresis loops it
is important  to measure  some of their  properties.  One of  the most
studied hysteretic  features is loop area. In fact,  it represents
the amount  of energy  dissipated during  a cycle and  thus it  is an
important quantity to be controlled  both from the theoretical and 
materials application point of view.   In Fig.~\ref{area} we show the
loop  area,  averaged  over  several  disorder  configurations,  as  a
function of the parameter $\lambda$, for various values of $\sigma$.

As can  be seen, the  area shows a  much more important  dependence 
on $\lambda$ for negative
than  for positive $\lambda$. This  behavior can
also be seen by studying the coercivity (amplitude) of the hysteresis
cycles at $m=0.5$, which displays a very similar dependence on $\lambda$.

In Figs.~\ref{lambda_vari}, \ref{sigma_vari} and \ref{rho_vari} we can
see that the hysteresis cycles  obtained with our model are asymmetric,
i.e. the decreasing field branch (transformation) cannot be related by
an    inversion   operation   to    the   increasing    field   branch
(retransformation).    This   is   an   interesting   property   since,
experimentally, materials displaying  a transition  to  a multivariant
phase show such behavior, which cannot be  reproduced with the RFIM
(see section \ref{Introduction}).  In  our model, this feature descends
from the  intrinsic difference of the physical  processes occurring in
the two branches  (transition from the $0$ state to  the three variant phase
in the first branch, and the  opposite process in the second branch).  In
order  to  study  this  feature more  quantitatively,  we  define  an
asymmetry factor $A$ as:
\begin{equation}
A=\frac{(dM/dH)_1-(dM/dH)_2}{(dM/dH)_1+(dM/dH)_2},
\end{equation} 
where  $(dM/dH)_1$   and  $(dM/dH)_2$  are  the   derivatives  of  the
hysteresis cycle at the coercive fields (defined as the two fields at
a height $m=0.5$) in the two branches.

\begin{figure}[ht]
\begin{center}
\includegraphics[width=7cm,angle=-90,clip=]{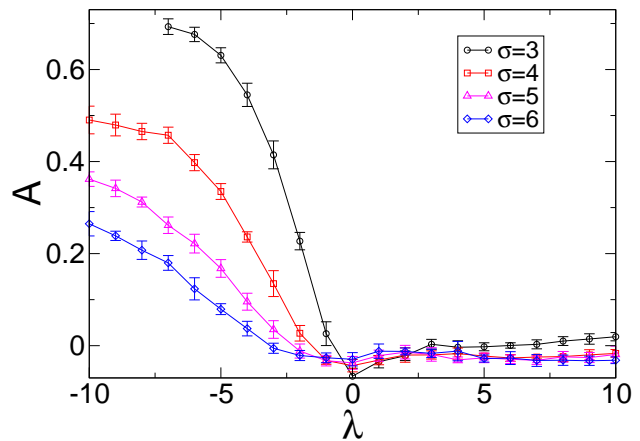}
\end{center}
\caption{\label{asimmetria} (Color on  line) Hysteresis loop asymmetry
  as a function of the parameters $\lambda$, for $\sigma=3,4,5,6$. The
  system  parameters are:  : $L=16$,  $\Delta H=0.05$ and
  $\rho=0$. Each  point represents an average  over $800$ realizations
  of the disorder. Solid lines are a guide to the eye.}
\end{figure}

As can  be seen in Fig.~\ref{asimmetria}, $A$  is greater than zero
for negative values of $\lambda$,  while for $\lambda>0$ the effect of
the 'dipolar' term is screened by  the disorder and the Potts term and
$A$ is essentially  $0$, irrespective of the value of  $\sigma$.  We have cut
the  curve with  $\sigma=3$ in  Fig.~\ref{asimmetria} at  the value
$\lambda=-7$.   In fact, for  the considered  range of  parameter, lower
$\lambda$  hysteresis  curves begin  to  show  the plateaux  explained
above, and thus our definition of asymmetry loses sense.

\begin{figure}[ht]
\begin{center}
\epsfig{file=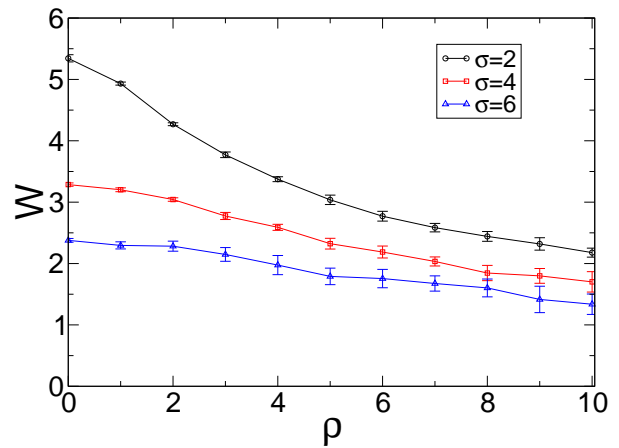,width=7.cm,angle=-90,clip=}
\end{center}
\caption{\label{larghezza} (Color on line)  Hysteresis loop width as a
  function  of the  parameters  $\rho$, for  $\sigma=2,4,6$. The  system
  parameters   are: $L=16$,  $\Delta   H=0.05$ and
  $\lambda=-3$.    Each  point  represents   an  average   over  $800$
  realizations of the disorder. Solid lines are a guide to the eye.}
\end{figure}

The  effect  of  disorder   on  hysteresis  loop properties  is
illustrated in  Fig.~\ref{larghezza}, where  we show the width $W$ of the
loops at  $m=0.5$ (coercivity) as a function  of increasing correlation
$\rho$,  for  different  values  of  $\sigma$.  As  mentioned  in  the
qualitative description  above, both $\rho$ and  $\sigma$ decrease the
width of the loops.

\subsection{Microstructures}
\label{Microstructures}

As we have already mentioned (see  section \ref{Model}), when  the dipolar
term is large enough compared  to the Potts term, the transformed domains
lose  their  spherical  symmetry  and  start  to  show  a  non-trivial
microstructure. The microstructure of a system is defined as the arrangement
of the variants of the product phase.

\begin{figure}[ht]
\begin{center}
\epsfig{file=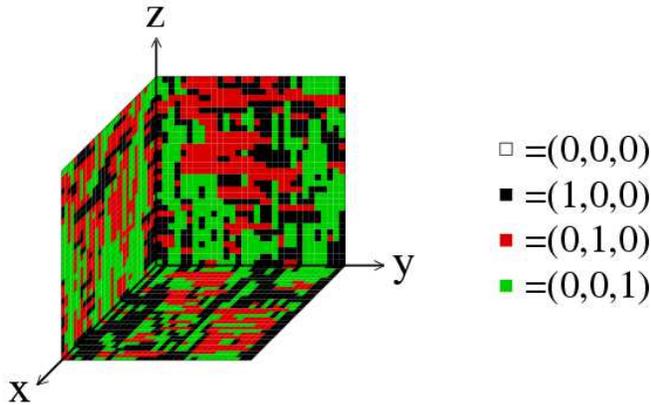,width=8.8cm,clip=}
\end{center}
\caption{\label{FIG1}  (Color on  line)  Saturation configuration  for
system parameters: $L=32$, $\Delta H=0.05$, $\lambda=-72$, $\sigma=20$
and  $\rho=20$.   Different   colors  correspond  to  different  spin
variants (see the legend).}
\end{figure}

In  Fig.~\ref{FIG1} we  can see an example  of these
three-dimensional microstructures.  
We  represent the views  of the $yz$, $xz$  and $xy$
surfaces,  when   the  sample   has  reached saturation  (fully
transformed state).
In  this case  ($\lambda<0$), the  domains  tend to  be prolate.   For
instance  green domains  (corresponding to  $\vec{S}_i=(0,0,1)$) tend  to be
elongated along  the $z$ direction both  in the $xz$  and $zy$ planes.
This effect can be quantitatively measured as will be explained below.

\begin{figure}[ht]
\begin{center}
\epsfig{file=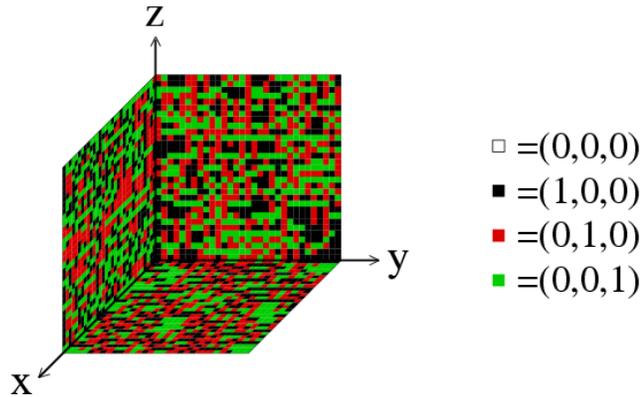,width=8.8cm,clip=}
\end{center}
\caption{\label{piani}  (Color on  line) Saturation  configuration for
system parameters:  $L=32$, $\Delta H=0.05$,  $\lambda=3$,
$\sigma=1$  and $\rho=1$.  Different  colors correspond to different
spin variants (see the legend).}
\end{figure}

For $\lambda>0$, we  observe the formation of oblate  domains, as shown
in Fig.~\ref{piani}, developing in the plane perpendicular to the spin
direction.  This effect generates a sort of ``chessboard'' correlation
as can be seen, for instance in the $yz$ plane by observing the red and
green domains.

In order to quantify the  shape of the domains in  such microstructures, 
we have
calculated  the  average linear  size  $\langle D_x\rangle$,  $\langle
D_y\rangle$  and $\langle  D_z\rangle$, of  the domains  of  the three
variants  $x$,  $y$  and  $z$,  along  the  three  spatial  directions
$\hat{x}$, $\hat{y}$  and $\hat{z}$, at  the saturation configuration.
For  instance,   the  average   size  matrix  corresponding   to  
Fig.~\ref{FIG1} and Fig.~\ref{piani}, are shown in Tables \ref{table1} and
\ref{table2}.   In  the  first   case  (corresponding  to  the  prolate
domains), the diagonal elements of the matrix are sensibly larger than
the others,  confirming the growth tendency of domains  along the
orientation of  each variant.   As is  quite obvious,  for symmetry
reasons, the  diagonal elements  can be averaged  giving what  we will
call  the  average linear  size  in  the  parallel direction  $\langle
D_{\parallel}  \rangle$  and the  off-diagonal  elements  can also be 
averaged giving the average linear size in the perpendicular direction
$\langle  D_{\perp}  \rangle$.   For  the  case  of  Fig.~\ref{FIG1}
($\lambda<0$) we get  $\langle D_{\parallel} \rangle=4.44\pm 0.20$ and
$\langle D_{\perp  }\rangle =  2.19 \pm 0.05$.   For the case of 
Fig.~\ref{piani} ($\lambda >0$), the tendency  of the system to form oblate
domains  is  confirmed by  the  values $\langle  D_{\parallel}\rangle=
1.175 \pm 0.005$ and $\langle D_{\perp}\rangle= 1.89 \pm 0.03$.

\begin{table}[ht]
 \begin{center}
\begin{tabular}{|c|c|c|c|}
\hline &$\langle D_x\rangle$&$\langle D_y\rangle$&$\langle D_z\rangle$
\\ \hline  $\hat{x}$& 4.35  $\pm$ 0.14 &2.21  $\pm$ 0.02 &  2.20 $\pm$
0.02\\ $\hat{y}$&2.04  $\pm$ 0.02&4.22  $\pm$ 0.13&2.15 $\pm$  0.02 \\
$\hat{z}$&2.27 $\pm$ 0.02& 2.28 $\pm$ 0.02&4.75 $\pm$ 0.07 \\ \hline
\end{tabular}
\caption{\label{table1}  Average   size  matrix  for   the  saturation
configuration of Fig.~\ref{FIG1}.}
\end{center}
\end{table}

\begin{table}[ht]
 \begin{center}
\begin{tabular}{|c|c|c|c|}
\hline &$\langle D_x\rangle$&$\langle D_y\rangle$&$\langle D_z\rangle$
\\  \hline $\hat{x}$&  1.172 $\pm$  0.003 &  1.927 $\pm$  0.014& 1.903
$\pm$ 0.014\\  $\hat{y}$& 1.862 $\pm$ 0.013& 1.173  $\pm$ 0.003& 1.868
$\pm$ 0.013 \\ $\hat{z}$& 1.876  $\pm$ 0.013& 1.887 $\pm$ 0.013& 1.181
$\pm$ 0.003 \\ \hline
\end{tabular}
\caption{\label{table2}  Average   size  matrix  for   the  saturation
configuration of Fig.~\ref{piani}.}
\end{center}
\end{table}

\begin{figure}[ht]
\begin{center}
\epsfig{file=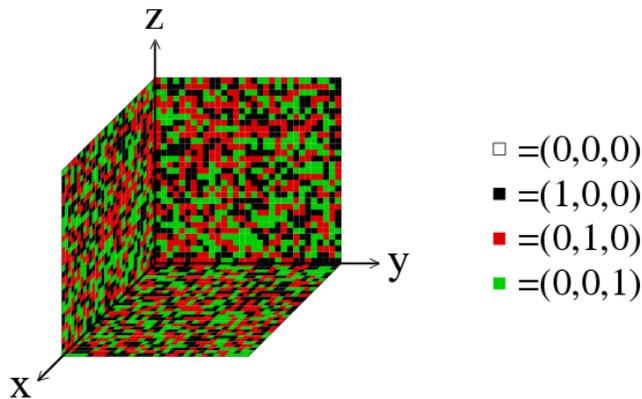,width=8.8cm,clip=}
\end{center}
\caption{\label{disordinoso}(Color  on line)  Saturation configuration
  for  system   parameters:  $L=32$,  $\Delta   H=0.05$,  $\lambda=0$,
  $\sigma=40$ and $\rho=40$. Different colors correspond to different
  spin variants (see the legend).}
\end{figure}

For the sake of completeness,  as a final microstructure   example,  in   
Fig.~\ref{disordinoso} we  show a  system configuration with  high disorder
and $\lambda=0$. As expected, no domain asymmetry arises and every spin
just  aligns with  its local  random  field. The  values of  $\langle
D_{\parallel}\rangle= 1.51 \pm 0.14$ and 
$\langle D_{\perp}\rangle= 1.50 \pm 0.20$ are
equal to within statistical errors.

\begin{table}[ht]
 \begin{center}
\begin{tabular}{|c|c|c|c|}
\hline &$\langle D_x\rangle$&$\langle D_y\rangle$&$\langle D_z\rangle$
\\ \hline  $\hat{x}$& 1.509  $\pm$ 0.082 &  1.491 $\pm$ 0.080  & 1.492
$\pm$ 0.079\\ $\hat{y}$& 1.517 $\pm$  0.083& 1.511 $\pm$ 0.081 & 1.504
$\pm$  0.080 \\ $\hat{z}$&  1.512 $\pm$  0.082 &  1.512 $\pm$  0.082 &
1.515 $\pm$ 0.081 \\ \hline
\end{tabular}
\caption{\label{table3}  Average   size  matrix  for   the  saturation
configuration of Fig.~\ref{disordinoso}.}
\end{center}
\end{table}

\begin{figure*}[ht]
\begin{center}
\epsfig{file=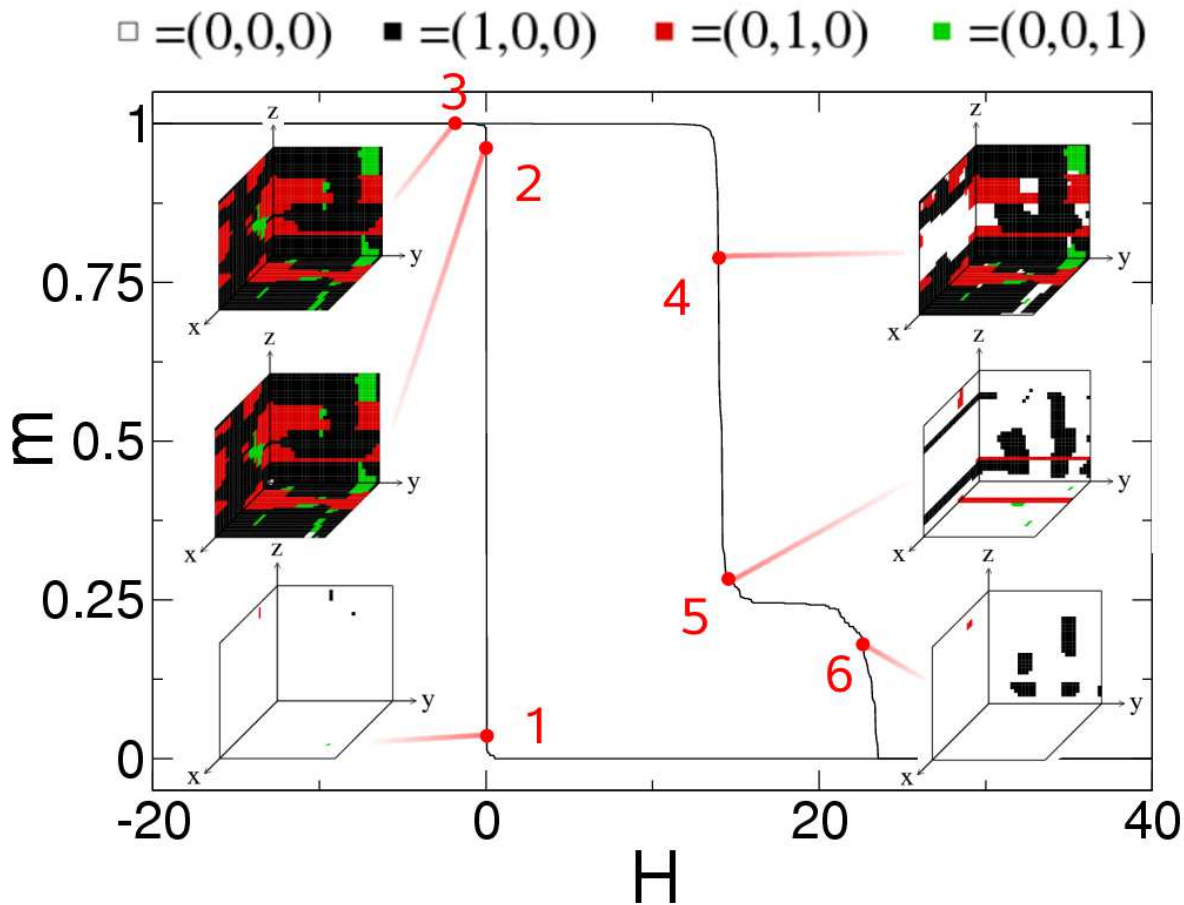,width=16.cm,clip=}
\end{center}
\caption{\label{figurone}(Color  on line) Parameters: $L=32$, $\lambda=-12$,
  $\sigma=1.5$, $\Delta H=0.05$ and $\rho=0$. The configuration snapshots are
  taken for (1) $m=0.03$; (2) $m=0.98$; (3) $m=1$; (4) $m=0.78$; (5) $m=0.28$
  and (6) $m=0.18$.}
\end{figure*}

The formation of microstructures such as those in Fig.~\ref{FIG1} and
\ref{piani} is  quite clearly affected by the  dynamics due to  the effect of
kinetic constraints. In  fact, when  a domain  of one
variant starts  to grow,  it necessarily blocks  the growth  of neighboring
domains of other variants and  {\it vice versa}. Thus the first variant
to locally break symmetry  will facilitate the  nucleation of large
domains, thus restricting  the dimensions  of  the  other variants.   This
effect could  be seen, for instance, by analyzing the decreasing field 
branch in  Fig.~\ref{figurone}: 
the formation of the domains of type $x$ and $y$ that cross the system
block the growth of domains of the $z$ variant.

Moreover, with the help of the microstructure representation, 
we can analyze in more detail the
bump formation due to finite size effects, discussed in section 
\ref{area_asimmetria} (see Fig.~\ref{figurone}): in fact, 
if the system size is finite there is
the formation of domains spanning a whole system side, such as the $y$ 
domains in microstructures labeled with $4$ and $5$ in Fig.~\ref{figurone}.
As already pointed out, 
these kind of slab domains are actually infinite due to the periodic boundary
conditions and are thus very stable. As we can see in the figure, they keep
existing up to high driving field, giving rise to the existence of the
plateau and disappear when the field overcomes a certain threshold.


As we already mentioned in  section \ref{Model}, the truncation of the
dipolar term  does not allow elastic effects to be reproduced, which would
lead  to more  realistic  microstructure. In  particular, it would  be
interesting  to control  the tendency  for the  different  variants to
exhibit a preferred  habit plane, as would be the case  of a real cubic
to  tetragonal  transition.   We   expect  that including a
next-nearest-neighbor dipolar  interaction  will  allow for  such  an
interesting property to occur. The model presented here is, therefore,
a promising starting point for modeling materials with a phase
transition from a single variant to a multivariant phase.

\subsection{Avalanches}
\label{Avalanches}

Another  phenomenon  that  may  be  analyzed with  our  model  is  the
avalanche  dynamics.   The hysteresis  loops  in athermal  first-order
phase transitions  consist of a  sequence of jumps  between metastable
states.   Such discontinuities  are, in general,  microscopic. However, for
certain values  of the disorder one  or more may  become comparable to
the  system  size and  then  correspond  to  the observed  macroscopic
discontinuities  in the  ferromagnetic loops.   In the  magnetic case,
microscopic avalanches  can be detected  experimentally by appropriate
coils.   They   correspond  to   the   so-called  Barkhausen   noise
\cite{Barkhausen1919}.  In  structural transitions, avalanches  can also
be    detected    typically    by   acoustic    emission    techniques
\cite{Vives1994a}.  Knowledge of the distribution of the number of
avalanches along the transition, as well as their size and duration, 
is an important
piece of information in order to characterize athermal FOPT.

Good discrimination of  individual avalanches in the simulations can
only be performed in the limit  of $\Delta H \rightarrow 0$. This will
require a  true adiabatic simulation algorithm which  is not available
at  present, as opposed to the case of the  standard  RFIM
\cite{Kuntz1999}.   In our  case, after  a small  (but  finite) change
$\Delta H$ some spins can be  updated. We will consider all of them as
being  part of  a single  avalanche.   This is  an approximation  and,
therefore, we should keep in  mind the fact 
that we are slightly overestimating the
size of the observed avalanches due to some unavoidable overlaps.
In the experiments recording avalanches the same problem occurs
\cite{White2003,PerezReche2004c}.

\begin{figure}[ht]
\begin{center}
\epsfig{file=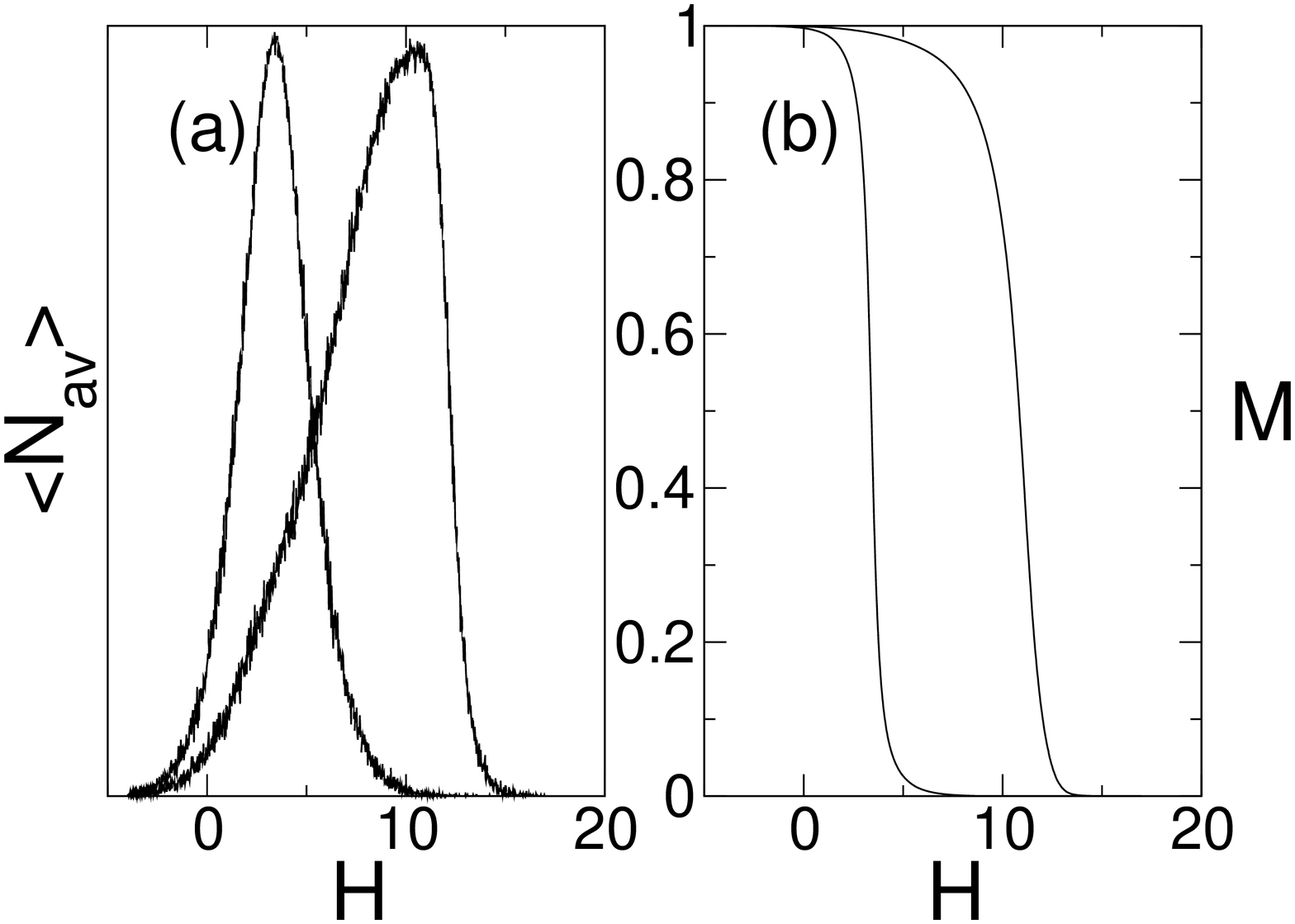,angle=-90,width=8.5cm,clip=}
\end{center}
\caption{\label{nvalanghe}   (a)   Number   of   avalanches   $N_{av}$
(arbitrary units)  as a function  of the external field.   $N_{av}$ is
computed in  each $\Delta H$ interval  over $600$ loops,  and (b): The
related  averaged  hysteresis  cycle. Simulation  parameters:  $L=16$,
$\Delta H=0.02$, $\lambda=-8$, $\sigma=4$, $\rho=0$.}
\end{figure}

With this definition we can study, for instance, the average number of
avalanches  $\langle N_{av}  \rangle$ as  a function  of  the external
field $H$.  As an example, in Fig.~\ref{nvalanghe}(a) we present the
behavior  of  this  number  for the  case  $\lambda=-8$,  $\sigma=4$,
$\rho=0$, compared  with the behavior of the  average hysteresis loop
in Fig.~\ref{nvalanghe} (b).  $\langle N_{av}  \rangle$ presents two
peaks in correspondence with the two coercive fields and goes to zero
(as  expected)  at the  $m=0$  and  $m=1$  saturations. This  kind  of
information   is  very   relevant  for   the  understanding   of  the
measurements of acoustic emission  in structural transitions using the
pulse-counting technique \cite{PerezReche2004}.

More  interesting  information  can   be  obtained  by  measuring  the
avalanche   sizes  $S$  and   computing  the   integrated  probability
distribution $P(S)$,  by analyzing all  the avalanches in a single branch of
the loop (the two branches must be analyzed separately 
since they are not symmetric). 
As a naive approximation, in our case 
one can define in our case the size $S$ of the
avalanche as the order  parameter variation $\Delta M$ associated with
an  avalanche  (i.e.   when  the  field  is  varied  by  $\Delta  H$.)
Nevertheless, this  definition imported from the  standard RFIM should
be carefully  adapted to our  multivariant FOPT.  Inside  an avalanche,
in fact, one can distinguish between different kinds of processes taking place,
depending on  their effect on  the order parameter variation.   Let us
focus on the decreasing field  branch starting from the $m=0$ phase up
to the  $m=1$ saturated  configuration: there are  several microscopic
possibilities for a  spin jump.  A spin could jump  from the $0$ state
to one of  the three variants $x$,  $y$ and $z$ thus giving  rise to a
positive contribution to the magnetization change $\Delta M$; it could
jump  from the  $x$, $y$  or  $z$ states  to $0$,  causing a  negative
contribution  $\Delta M<0$;  or finally  there is  a possibility  of a
change from one variant to  another without contributing to the change
in  the system magnetization.   These three  possible updates  will be
called  $+$, $-$ and  $0$. Instead  of only measuring the  total size
$\Delta M$ of an avalanche, for each of them we will measure the three
quantities  $n_+$, $n_-$  and  $n_0$,  which are  the  number of  spin
updates  of each kind.   Moreover, since as we  have seen  in section
\ref{area_asimmetria},  the  hysteresis loops  are  not symmetric,  we
should separately analyze the avalanches during  the decreasing field
branch and during the  increasing field branch. This gives, therefore,
6    possible    distributions:   $P_+^{down}(n)$,    $P_-^{down}(n)$,
$P_0^{down}(n)$  for the  decreasing field  branch  and $P_+^{up}(n)$,
$P_-^{up}(n)$,  $P_0^{up}(n)$  for the  increasing  field branch.   It
should be  mentioned that in  the increasing field branch the total
number of  events of the  $+$ and $0$  kind are much smaller  than the
number of  $-$ events, typically by  2-3 orders of  magnitude. For the
decreasing branch the $-$ events are  rare, but $0$ events  are frequent,
since  an important number of transitions  within variants  may occur
during the avalanches in the later stages of the transition.  Of course
such  an optimization  between variants  cannot occur  in  the reverse
process during the increasing field branch.

\begin{figure}[ht]
\hspace{4cm}
\centerline
\centerline{ \epsfig{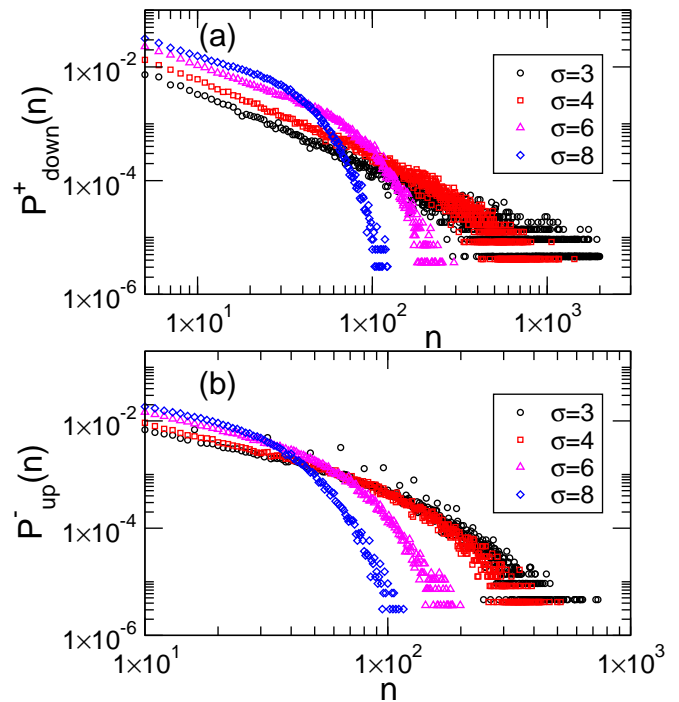}}
\caption{\label{PdiS_sigmavar}(Color  on  line)  Avalanche size  probability  distributions
$P_{+}^{down}(n)$ (a) and   $P_{-}^{up}(n)$ (b) corresponding  to   the  two
branches  of  the  hysteresis   loop,  averaged  over  $600$  disorder
realizations  and for  four values  of  $\sigma$ as  indicated by  the
legend.   Simulation   parameters  are  :   $L=16$,  $\Delta  H=0.05$,
$\lambda=-8$ and $\rho=0$.}
\end{figure}

\begin{figure}[ht]
\hspace{4cm}
\centerline
\centerline{ \epsfig{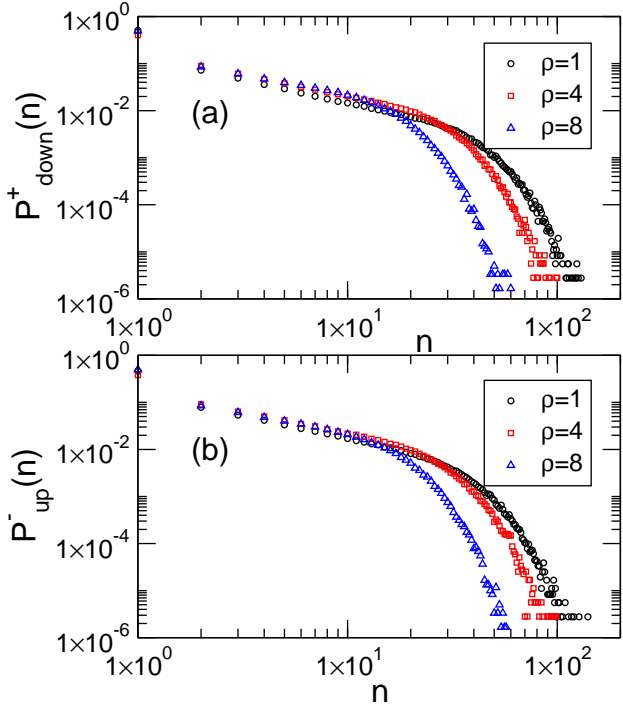}}
\caption{\label{PdiSrho}  (Color  online)  Avalanche size  probability
distributions  $P_{+}^{down}(n)$ and $P_{-}^{up}(n)$  corresponding to
the two branches of the  hysteresis loop, averaged over $600$ disorder
realizations and  for three  valued of $\rho$.   Simulation parameters
are: $L=16$, $\Delta H=0.05$, $\lambda=-8$ and $\sigma=8$.}
\end{figure}

In  Figs.~\ref{PdiS_sigmavar}  and  \ref{PdiSrho} we  show some  examples 
of  the  probability  distributions for  varying
values  of  the two  disorder  parameters on  log-log
plots,  respectively $\sigma$  and
$\rho$.   Actually,  we  show  only  the   $P_{+}^{down}(n)$  and  the
$P_{-}^{up}(n)$ distributions  which account  for the majority  of the
events. In Fig.~\ref{PdiS_sigmavar}(b) it is possible to 
notice another finite size effect: the slab domains of size
corresponding to multiples of the system size $L$ present the tendency
to disappear abruptly (and thus the $P^-_{up}(n)$ presents some peaks in
correspondence of values multiple of $L$), 
as we have already pointed out in section \ref{Microstructures}.

An   interesting  result   for  the   case  of   $\rho=0$   (see  
Fig.~\ref{PdiS_sigmavar}(a))  is  that  the distribution  $P_{+}^{down}(n)$
shows a exponentially damped behavior but seems to become a power-law
for  a certain  critical  value  of $\sigma$  (which  will be  located
around $\sigma \simeq  3$). Such a tendency has  been well studied in
the  standard RFIM.   The fitted  value of  the power-law  exponent is
1.1, clearly 
smaller    than     the    expected    universal     value    $2.03\pm
0.03$ \cite{Perkovic1999}, but this feature could be due to 
finite size effect. Only
a  detailed finite-size  scaling analysis  and the study of the geometry of
the
avalanches \cite{PerezReche2003} (out of  the scope  of this
paper) will reveal if the observed power law conforms to universal
behavior  or not.   Interestingly it  seems that  for  the increasing
field branch  (see Fig.~\ref{PdiS_sigmavar}(b)),  the distribution is
always exponentially damped, at least  for the studied range of values
of  $\sigma$. Therefore, the  critical point  for the  increasing field
branch, if  it exists,  would be located  at a different  (smaller)
value of $\sigma$.

\section{Summary}
\label{Conclusions}

The analysis of microstructure formation in ferroic systems undergoing
a first-order  phase transitions is  an interesting issue both  from a
purely theoretical  and an applicative point  of view. Microstructures
arise  since the  product phase,  arising  from the  balance of  many
energetic terms,  may show energetically  equivalent variants. Despite
the interest in this issue, the models  that have been used  up to now
for the study of the interplay between disorder and athermal evolution
(for example, the Random  Field Ising model \cite{Sethna1993}) are not
suitable  for the analysis  of microstructure  formation, due  to the
equivalence of the variants of the product phase.

In the present  work, we have introduced a  modification of the Random
Field Potts model, which consists of adding a dipolar term truncated to
the nearest-neighbor approximation,  which represents a promising step
towards the  analysis of athermal  transitions from a degenerate  to a
multivariant   phase.   

In our  simulations we have chosen extremal updating in order to
preserve  the independence of  the trajectory  from the  applied field
rate.  We have studied the dependence of the hysteresis loop shape on
the hamiltonian parameters values.  From a quantitative point of view,
this has been performed by measuring  the the loop area, its 
asymmetry and width. 
Our loops display a large area and asymmetry regime for very
negative values of the  'dipolar' term parameter $\lambda$, associated
with the formation  of microstructures  with prolate  domains, oriented
along  three equivalent spatial  directions.  On  the other  hand, for
$\lambda>0$  domains are  planar (oblate) and  the  loops show  low, almost
constant values of the area and the asymmetry.
 
We  have  also addressed the  study of  the  avalanches  in  the
hysteresis  loop, distinguishing  between the two branches.   For certain
parameter  values,  the  probability  distribution of the size of the leading 
process in an avalanche 
displays power-law behavior, which is the typical result for athermal
phase transitions in disordered systems.

\acknowledgments The authors  acknowledge fruitful discussions with A.
Planes.  This work has  received financial support from CICyT (Spain),
project MAT2007-61200  and CIRIT (Catalonia),  project 2005SGR00969 and
Marie Curie RTN MULTIMAT (EU), Contract No. MRTN-CT-2004-5052226.

\end{document}